\documentclass{article}

\usepackage{graphicx}
\date{}

\begin{document}

\title{Comment on the vortices of the early phase of the ELMs}
\author{F. Spineanu and M. Vlad  \\
National Institute of Laser, Plasma and Radiation Physics \\
Magurele, 077125 Bucharest, Romania}
\maketitle

\begin{abstract}
We discuss a possible perspective on the process of transition
from the layer of sheared poloidal rotation in the H-mode to a set of discrete filaments
that concentrate vorticity and current density. They may be precursors of the Edge Localized Modes.
\end{abstract}

\bigskip 

The transport barrier that exists in the H mode regime in tokamak has
frequently been identified in experimental observations with a layer of
sheared poloidal rotation. There are theoretical basis for this. It has been
shown that, in the presence of sheared poloidal rotation, the linear
eigenmodes are shifted relative to the resonant surface such that their
ability to extract the free energy from background gradients (\emph{i.e.} $%
\gamma $) is diminished. Moreover, the sheared poloidal flow tears apart the
radially elongated eddies of the drift wave instability and reduces the
radial correlation length of the drift wave turbulence. Since the diffusion
coefficient depends on the square of the radial extension of fluctuations it
results a substantial reduction of the transport. A transport barrier is
generated, in a narrow layer limited by the Last Closed Flux Surface. 

The origin of the sheared poloidal rotation is still unclear. The loss of
fast ions from the NBI, the Reynolds stress transfered by turbulent fields,
the particular dependence of the parallel viscosity on rotation are possible
sources; or of a radial (polarization) electric field \cite{Burrell}. The
growth of the poloidal rotation has two difficulties to overcome: one is the
inertia factor, approximatively $\left( 1+2q^{2}\right) $ and the other is
the damping due to the magnetic pumping. Then, if the transport barrier
consists of a sheared poloidal rotation, there should be a drive with
sufficient magnitude to create and then to maintain it.

In experiments it is systematically observed toroidal rotation and there is
a natural tendency to look for a connection with the enhanced regimes.
However the sheared poloidal rotation is much more efficient in suppressing
the radial transport than the sheared toroidal rotation. On the contrary,
the sheared toroidal rotation can be a source of turbulence. We will
consider here that the transport barrier is due to sheared poloidal
variation.

The shear of the poloidal velocity is the dependence of $v_{\theta }$ on the
radial coordinate, $\partial v_{\theta }/\partial r$. At the edge this is
the dominant term in%
\[
\left. \mathbf{\nabla \times v}\right\vert _{z}=\frac{1}{r}\frac{\partial }{%
\partial r}\left( rv_{\theta }\right) =\frac{v_{\theta }}{r}+\frac{\partial
v_{\theta }}{\partial r}
\]%
where $z$ is the toroidal direction. This means that the velocity shear is
the $z$ vorticty $\omega _{z}\approx \frac{\partial v_{\theta }}{\partial r}$%
. The formulation in terms of vorticity provides a wide range of
connections. In general the vorticity is a field that has a tendency to
self-organization \cite{FlorinMadi2003}, \cite{FlorinMadiXXX}. In a $2D$
fluid there is attraction between like-sign vortical elements, there are
vortex coalescences into larger and larger vortices. There is also
separation and clusterization of vorticities of opposite signs in different
regions of the plane. These have been observed in experiments on neutral
(Euler) fluids and have been explained theoretically. It has also been
observed in non-neutral plasma and in numerical simulations of $2D$ plasma
in strong magnetic field. Without intending to transport all this body of
knowledge to tokamak plasma, we can be guided in the exploration of H mode
and of Edge Localized Modes (ELM).

Adopting the vorticity as a basic analytical instrument of description, we
can refer to some known results from MHD. In numerical simulations of $2D$
MHD it has been found that the concentration of vorticity and the
concentration of the current density occur together and that asymptotically
the extremum of these two fields coincide. A monopolar vortex that
concentrates the initial vorticity field coincides with the maximum of the
current density. This already should be seen as a suggestion that, in the H
mode, the narrow layer of sheared poloidal rotation (\emph{i.e.} a layer of
vorticity) must also be a layer of concentration of the current density.
This must be seen as independent of the connection which is usually made
between the strong pressure gradient (due to the transport barrier) and the
bootstrap current. The current in the layer can be different of the
bootstrap value since the accumulation of current density has also a
dependence on local vorticity. Numerical simulations of $2D$ MHD have
clearly shown that, if initialized in separate regions on the plane, the
current and the vorticity evolve to become overlapped and further they do
not separate anymore. 

\bigskip 

We now have to examine the situation of a layer of vorticity and current,
under strong transversal gradients of density and temperature. The barrier
of transport exists due to this sheared rotation but this has only
suppressed the convective transport that is usually realized by turbulent
eddies on finite radial distances. The conductive transport remain possible
although it is much smaller. We are here in a configuration analogous to the
Rayleigh Benard experiment. As in that case, if the gradient of
temperature/pressure increases the convective plumes that always exist from
the internal surface toward the external limit of ths layer begin to
organize themselves into large scale convective pattern, the cells. In our
case this means formation of filaments almost parallel to the magnetic field
line) which break up the rotation layer (and implicitely the barrier of
transport). The filaments consists of vorticity and current density. This
phase should be seen as the basis for the further evolution according to the 
\emph{peeling balloonning} instability. 

The transition from a layer of sheared rotation to a discrete set of
vortices is a well known phenomenon in various areas. 

In the physics of atmosphere, the ring-type shape of vorticity distribution
in a tropical cyclone is sometimes broken and transformed into a set of
vortices placed symmetrically on the circle \cite{Kossin}. In non-neutral
plasma this is observed for the electron density distribution. In fluid
systems, the Kelvin Helmholtz instability creates in a layer of sheared flow
discrete centers of vorticity concentration that grow and finally break up
the layer. 

To the vorticity concentration we must add the tendency of the layer of
current density (practically the transport barrier is a current sheet) to
generate concentration in a set of filaments of current.

\bigskip 

The universal prototype for transition from a sheared layer to a set of
concentrated filaments of vorticity and current is the nonlinear evolution
of a Chaplygin gas with anomalous polytropic. It has been examined in detail
and illustrated with many examples by Trubnikov \cite{Trubnikov}.
One of the applications is the break up of a layer of density and of current
in a set of filaments \cite{BulanovSasorov}. 

The geometry adopted by Trubnikov is adequate for studying the tearing of
the density distribution in the layer. The width (along $y$) is uniform
initially $L_{0}$ and it evolves to a profile $L$ which is variable along
the direction $x$ of the layer. The coordinate $y$ is perpendicular on the
layer in the equilibrium position. We can see this as $y$  \emph{radial} and 
$x$  \emph{poloidal}.

The magnetic field has a shear $B=B_{x}\left( y\right) =-B_{0}\tanh \left( 
\frac{y}{L}\right) $ where $B_{0}$ is the main magnetic field, along $z$.
The current density is%
\[
j_{z}=en\left( v_{iz}-v_{ez}\right) 
\]%
Then%
\[
v_{iz}-v_{ez}=\frac{cB_{0}}{2\pi enL\left( t,x\right) }=\frac{const}{%
nL}
\]%
It is introduced a normalized density of plasma $\rho \left( t,x\right) =%
\frac{nL\left( t,x\right) }{n_{0}L_{0}}$ and we have the usual density
conservation%
\[
\frac{\partial \rho }{\partial t}+\frac{\partial }{\partial x}\left( \rho
v\right) =0
\]

Under the assumption $v_{e,th}<v_{ez}$ we have the equation of motion%
\[
\frac{\partial v}{\partial t}+v\frac{\partial v}{\partial x}=\frac{1}{nm_{i}c%
}\left( -j_{z}B_{y}\right) =\frac{e}{m_{i}c}\left( v_{iz}-v_{ez}\right) 
\frac{\partial A}{\partial x}
\]

We consider that the system is invariant along the $z$ direction which means
that the generalized momenta of the electrons and of ions are conserved $%
m_{i}v_{iz}+\frac{e}{c}A=$const and $m_{e}v_{ez}-\frac{e}{c}A=$const$%
^{\prime }$ . The equation of motion becomes%
\[
\frac{\partial v}{\partial t}+v\frac{\partial v}{\partial x}=\frac{e}{m_{i}c}%
\left( v_{iz}-v_{ez}\right) \frac{\partial A}{\partial x}=c_{0}^{2}\frac{1}{%
\rho ^{3}}\frac{\partial \rho }{\partial x}
\]%
The constant is $c_{0}^{2}$ is a constant. The two equations are%
\begin{eqnarray*}
\frac{\partial \rho }{\partial t}+\frac{\partial }{\partial x}\left( \rho
v\right)  &=&0 \\
\frac{\partial v}{\partial t}+v\frac{\partial v}{\partial x} &=&c_{0}^{2}%
\frac{1}{\rho ^{3}}\frac{\partial \rho }{\partial x}
\end{eqnarray*}%
Obviously this is like a "gas with anomalous polytropic" , which means that
a perturbation accumulates more and more gas, with formation of separated
maxima. The equations are solved using a \emph{hodograph }transformation.
The solution is determined by \textbf{Trubnikov} and consists of a set of
maxima (corresponding to filaments) that result from breaking the initially
uniform distribution of density in the layer of current. The solution shows
the dynamical process of replacement of the layer with the discrete
filaments.

\vspace*{0.1cm}
\begin{center}
\begin{figure}
    \centering
    \includegraphics[height=7cm]{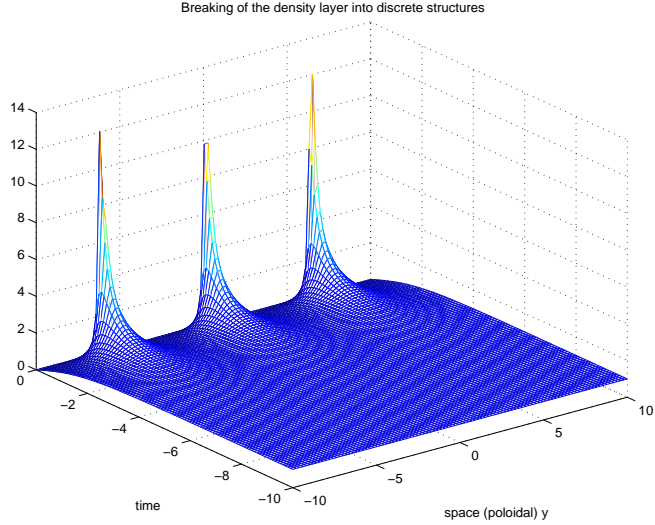}

\caption{\small Trubnikov solution showing the filamentation as function of time.}
\end{figure}
\end{center}

This is a useful example where the H mode barrier, which is a layer of
vorticity and current density, can be broken and replaced by discrete
filaments. It would be interesting to see how this dynamical process evolve
to become the peeling-balloonning instability. It is suggested that the
early phase of the ELM consists of the formation of filaments or vortices
where the vorticity and the current density are concentrated. 

\bigskip 

\textbf{Note.} This is part of a work presented at the meeting "EFDA-TG
Transport. 2nd meeting JET Culham 16 - 18 September 2009"

\end{document}